\documentclass[American paper,twocolumn]{esapub} 
\usepackage{graphicx}
\usepackage{natbib}

\title{Population synthesis of neutron stars, strange (quark) stars and
black holes }
\author{K. Belczynski${^{1,2}}$}
\author{T. Bulik$^1$}
\author{W. Klu{\'z}niak$^1$}

\affil{$^1$Nicolaus Copernicus Astronomical Center, Bartycka 18,
00-716 Warsaw, Poland}
\affil{ $^2$Harvard-Smithsonian Center for Astrophysics, 60 Garden
St., MS. 51, Cambridge, MA 02138, USA }

\begin{document}

\maketitle

\begin{center}
Abstract
\end{center}
\vspace*{-0.5cm} 
We compute and present the distribution in mass
of single and binary neutron stars, strange stars, and black
holes. The calculations were performed using a stellar
population synthesis code. We follow all phases of single and
binary evolution, starting from a ZAMS binary and ending in the
creation of one compact object (neutron star, black hole,
strange star) and a white dwarf, or two compact objects (single
or binary). We assume that neutron stars are formed in the
collapse of iron/nickel cores in the mass range $M_0 < M < M_1$,
quark stars in the range $M_1 < M < M_2$, and black holes for
core masses $M>M_2$ and find that the  population of quark stars
can easily be as large as the population of black holes, even if
there is only a small mass window for their formation\vspace*{0.1cm}.

Key words: compact objects -- masses, binaries -- population 
synthesis\vspace*{1cm}.

\section{Introduction}

Binary population synthesis is a useful tool for studying  the
statistical properties of stars, including the compact objects.
In this  paper we wish to address the following questions: what
is the distribution of masses of the compact objects? what is
the relation between the stellar initial mass and the final mass
of a compact object, when binary evolution is taken into
account? given the distribution of compact object masses, what
are the relative numbers of different types of objects (neutron
stars, quark stars, black holes)?

In section 2 we shortly describe the population synthesis code
used here, in section 2 we present the results, and finally in
section 4 we present the conclusions.

\begin{figure*}
\includegraphics[angle=-90,width=0.95\textwidth]{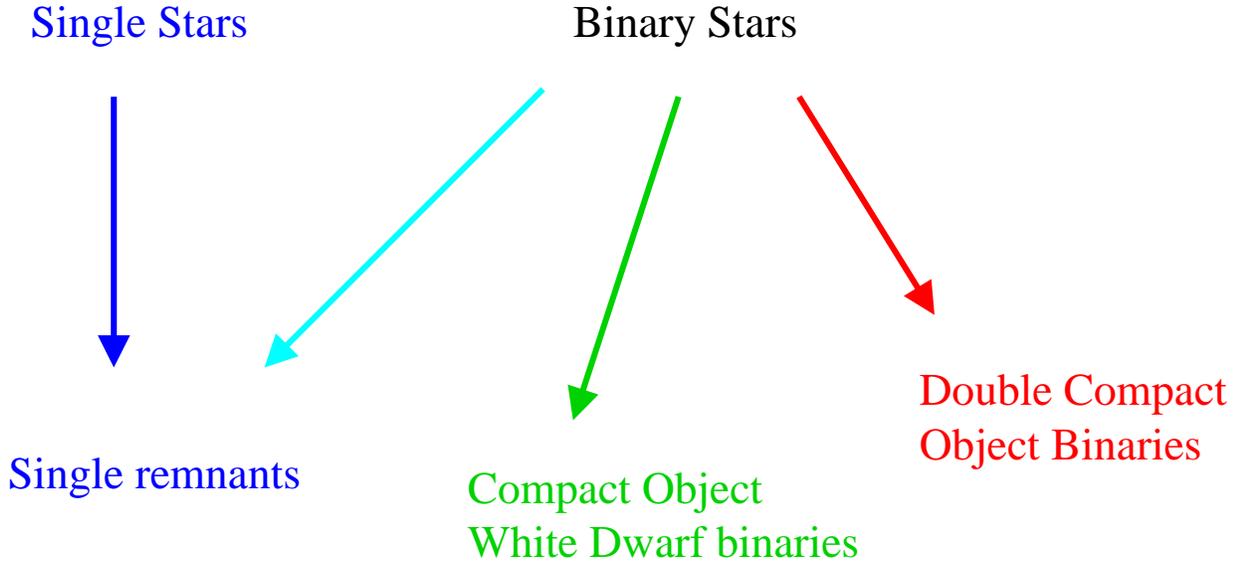}
\caption{The various routes of compact object formation
(compact object = a neutron star, a strange star or a black hole).
}
\end{figure*}

\section{Population synthesis code}

We use the stellar binary population synthesis code  consisting
of two parts. The single star evolution is based on modified
formulae from Hurley et al. (2000).  We have changed the
prescription for mass of the compact object formed in a
supernova explosion. We use original Hurley et al. (2000)
formulae to obtain final CO  core mass. We use models of Woosley
(1986) to calculate the final  FeNi core mass (for a given CO
core mass), which will collapse and form a compact  object
during supernova explosion. Finally, we include calculations of
Fryer and Kalogera (2000) to  take into account black hole
formation both through fall back and directly. 

The binary evolution is described in Belczy{\'n}ski et al. (2000). 
We evolve only binaries where at least one star  will  undergo 
a supernova explosion. 
The evolution starts at ZAMS. During the course of evolution 
we include the following effects:
{ \begin{itemize}
 
             \item wind mass loss (standard, Wolf-Rayet, LBV)
             \item tidal circularization of binary orbit
             \item magnetic breaking          
             \item conservative/nonconservative mass transfer
             \item common envelope evolution
             \item rejuvenation
             \item hyper-accretion onto compact objects
             \item detailed supernova explosion (SN) treatment
             \item partial and complete fall back onto compact objects in SN             
 \end{itemize}
}
\begin{figure}
\includegraphics[width=0.95\columnwidth]{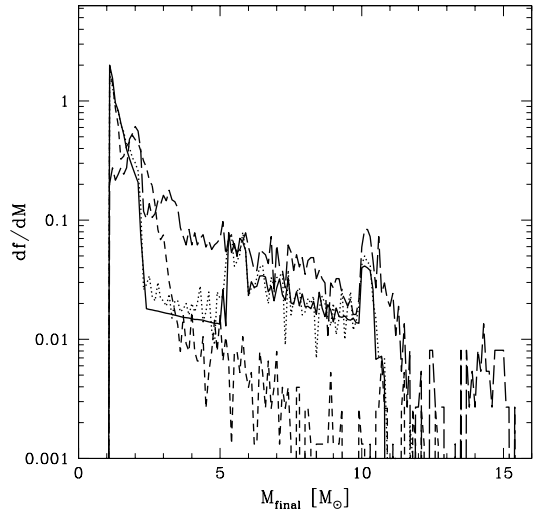}
\caption{The number per mass interval of neutron
stars / strange stars / black holes formed along each route shown 
in Fig.1. The solid line corresponds to the case of single
stellar evolution, the dotted line represents the single compact
objects formed in binary evolution (group I), the short dashed
line shows the compact objects in white dwarf binaries (group II), and the
long dashed line - the compact objects in double compact object
binaries (group III).
}
\end{figure}  
        
Many binaries are disrupted in supernova explosions, as a result
of mass loss and the natal kick. For supernova kicks we use
distribution presented by Cordes and Chernoff (1997). We
continue to evolve all stars, both single and binary components,
until the  formation of a  stellar remnant.

\section{Results}

In { Figure 1} we show different formation routes of compact 
objects: neutron stars, strange stars and black holes. Compact
objects are formed both through  single and binary stellar
evolution. Single compact objects are  descendants of massive
single stars but may  also be formed as a binary system is
disrupted in a supernova explosion of  one of its components
(these are the paths on the left hand side of   Fig.1).  We will
denote the single compact objects formed as a result of the
binary evolution as group I. Under favorable conditions some
binaries survive supernova explosions, and they finally form
tight systems with compact object/objects. Most of these
binaries will consist of a white dwarf and a compact object and
the rest will form binaries with two compact objects  (we will
denote the compact object in binaries with white dwarfs as group
II, and the double compact objects as group III).

In Figure 2 and 3 we show production rates of compact objects as
a function of their final mass for the  four different formation
routes presented in Fig.1.

\begin{figure*}[t]
\centerline{\includegraphics[width=0.8\textwidth]{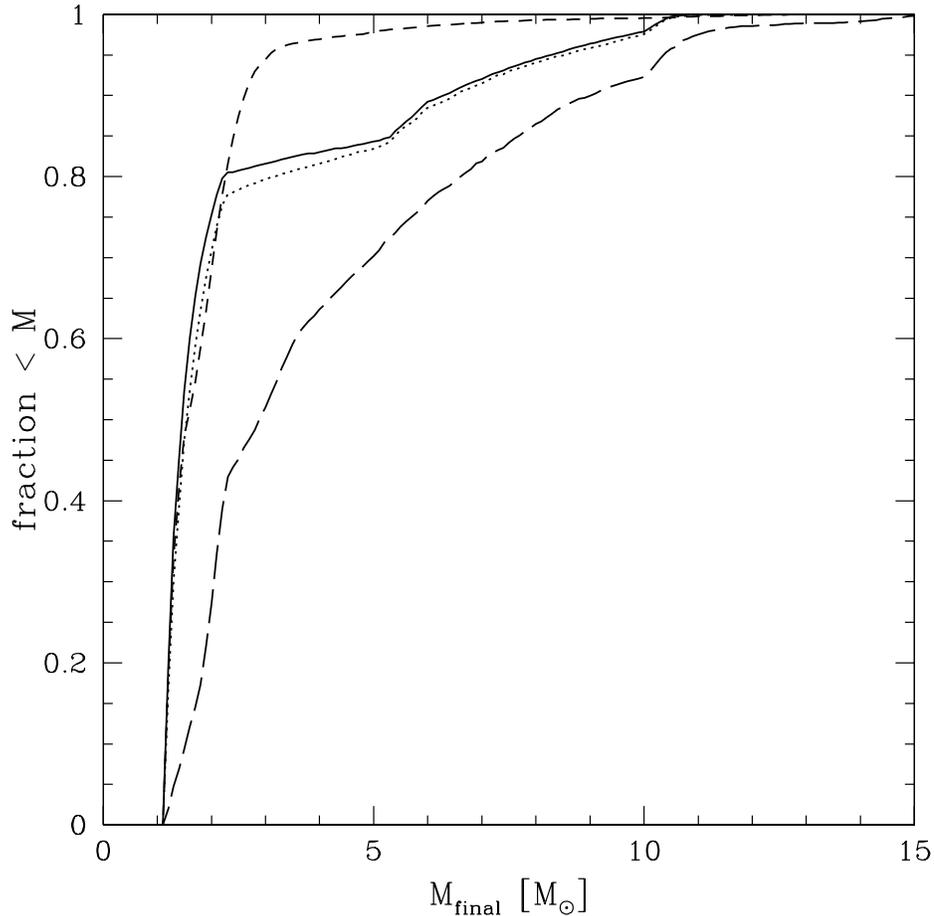}}
\caption{ The cumulative fraction of comapct objects (neutron
stars, strange
stars, black holes) corresponding to the differential distribution of Fig.2.
The line types are the same as in Figure 2.}
\end{figure*}

{ Figure 2} shows the number of compact objects  per mass
interval  formed along each route of Figure 1. We start forming
compact objects at mass $\sim 1.2 M_\odot$ and their number 
falls of with the mass of a final compact object, as expected
for our assumed  initial mass function $\sim M^{-2.7}$. A peak
around $\sim 10 M_\odot$ reflects our relation between ZAMS mass
of a progenitor and the final mass of a compact object, which is
shown for single stars with a solid line in Fig.4. This relation
for wide range of progenitor ZAMS masses results in a final
compact object mass of $\sim 10 M_\odot$. This is an effect of a
stellar wind which increases with a stellar mass, and  thus
decreases  the final mass of a compact object flattening out the
relation  at about $10 M_\odot$. 

{ Figure 3} shows cumulative fraction of compact objects as a
function of their final mass for different formation routes.
Note the very fast initial rise of cumulative curves for single
compact remnants and  remnants  white dwarf binaries. 
Most of compact objects in these groups have mass below $2
M_\odot$. Only for double compact objects a large fraction
of remnants have  higher final masses.  

In { Figure 4} we plotted the mass of a single compact object as
a  function of the ZAMS mass of a progenitor star. Single
compact objects coming from single stars are shown with solid
line,  and the ones coming from disrupted binary systems are
shown with points. Points which are over the solid line form a
group of compact objects whose  progenitors gained mass in
binary interactions, and thus they formed more massive compact
objects then corresponding single stars of the same ZAMS mass.
Points which are below the solid line represent a group of
compact objects whose  progenitors were donors in different mass
transfer events, and thus they lost part of their mass and 
formed less massive compact objects. There is also a group of
compact objects coming from binary progenitors, which  follows
exactly M$_{\rm zams}$--M$_{\rm final}$ relation for compact
objects formed from single stars. This group comes from wide,
non interacting binaries, which in the most cases were disrupted
in a first supernova explosion. The exploding component was
unaffected by its companion so it has left behind a compact
object remnant exactly as it would be formed from a single
star.    Then its companion, also unaffected by binary
evolution, continues its evolution.  Provided that it is massive
enough, it goes through a supernova explosion  leaving second
single compact object remnant, of a mass fitting the single 
star M$_{\rm zams}$--M$_{\rm final}$\ relation.

\begin{figure}                    
\includegraphics[width=0.9\columnwidth]{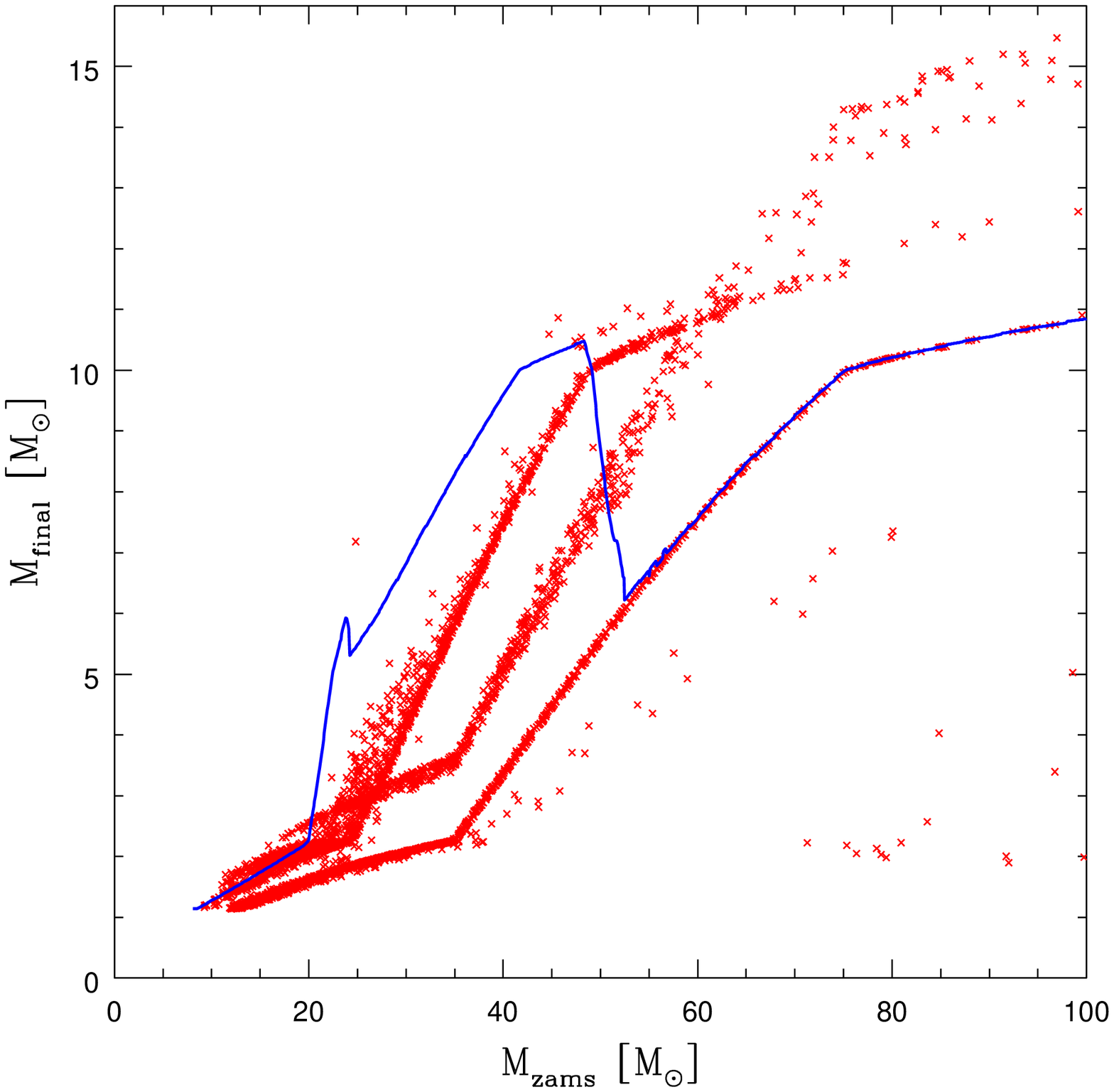}
\vspace{-2.5mm}

\includegraphics[width=0.9\columnwidth]{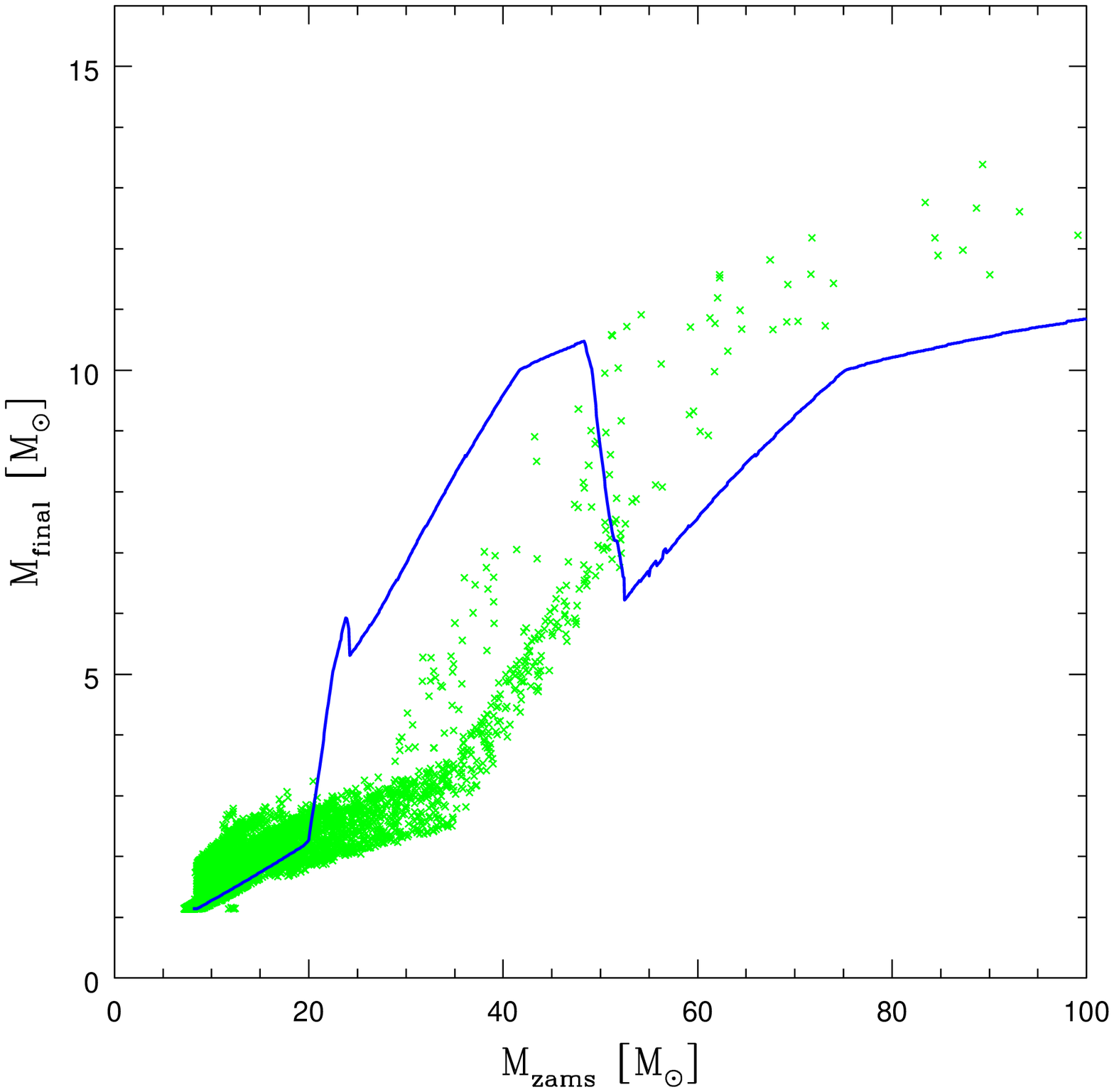}
\vspace{-2.5mm}

\includegraphics[width=0.9\columnwidth]{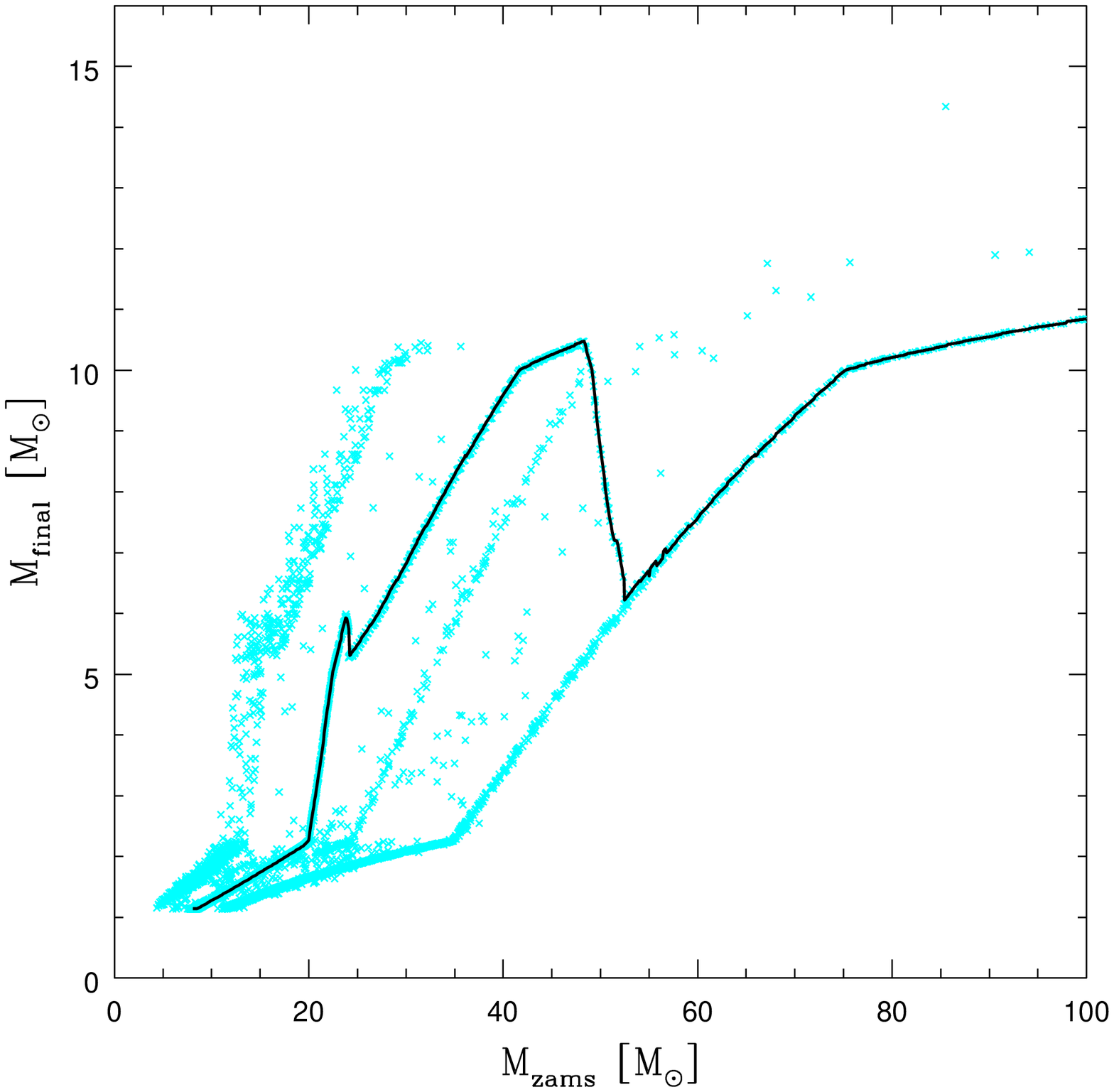}
\vspace{-2.5mm}

\caption{The mass of a single compact object as a function  of
the ZAMS mass of the progenitor star. The solid line shows the 
 correspondence between the initial and final mass for
initially single stars. The points correspond to  compact
objects whose progenitors were initially in binary systems.
The three panels correspond to group I, II, and III, from to 
bottom respectively.
 Note that binary stars can have more then one fate.}

\end{figure}

\section{Conclusions}

We have shown the effects of the binary evolution on the
distribution of masses of compact objects. As expected the bulk
of the population of compact objects have masses below
$2\,M_\odot$. While for single stellar evolution  there exists a
unique relation between the stellar mass and  the compact object
mass,  there is no such relation when the binary evolution is
taken into account. The binary evolution works both ways,
the masses of the compact objects formed in binaries can be both
smaller and larger than in the corresponding case of single
stellar evolution.

The masses of the compact objects are concentrated in the low
mass range - see Figure 3. Thus even if the mass interval
($M_1,M_2$) for the formation  of strange (quark) stars is
small,  the fraction of strange stars in the whole compact object
population can be huge. The exact fraction as a function  of
$M_1$ and $M_2$ can be easily read off Figure 3.  On the other
hand the fraction of black holes hardly depends on $M_2$ (the
cumulative curves flatten above $3\,M_\odot$. We conclude that
the population of quark stars can easily be as large as the
population of black holes, even if there is only a small mass
window for their formation\vspace*{0.5cm}.

\begin{center}
Acknowledgments
\end{center}
\vspace*{-0.5cm}
The authors gratefully acknowledge support of the 
KBN grant 2P03D00418\vspace*{0.5cm}.


\begin{center}
References
\end{center} 
\vspace*{-0.5cm}

Hurley J.R., Pols O.R., Tout C.A., 2000, MNRAS 315, \vspace*{0.2cm}543\\
Belczynski K., Kalogera V., Bulik T., 2000, in \vspace*{0.2cm}preparation\\
Cordes J., Chernoff D. F, 1997, ApJ, 482, \vspace*{0.2cm}971\\
Woosley S.E., 1986, in 'Nucleosynthesis and Chemical Evolution', 
16th Saas-Fee Course, eds. B. Hauck et al., Geneva Observatory, 
\vspace*{0.2cm}p.1\\
Fryer C.L., Kalogera V., 2000, astro-ph/9911312\\

\end{document}